\documentclass[aps,prl,showpacs,superscriptaddress,reprint]{revtex4-2}
\usepackage{rotating}
\usepackage{epsfig} 
\usepackage{xcolor}
\usepackage{url}

\usepackage{multirow}
\usepackage{makecell}

\usepackage{amsmath}
\usepackage{graphicx}

\usepackage{lineno}

\usepackage{hyperref}

\graphicspath{ {./plot/}, {./image/} }
\DeclareGraphicsExtensions{ .pdf, .png}

\begin{document}

%------------------------------------------------------------------------------------------------%

\newcommand{\der}{\text{d}}
\newcommand{\rp}{\Psi_{\textsc{rp}}}
\newcommand{\pt}{p_{T}}
\newcommand{\gevc}{GeV/$c$}
\newcommand{\Ru}{{\rm Ru}}
\newcommand{\Zr}{{\rm Zr}}
\newcommand{\Ntrk}{N_{\rm trk}}

\newcommand{\qcd}{{\textsc{qcd}}}
\newcommand{\pos}{{\textsc{os}}}
\newcommand{\pss}{{\textsc{ss}}}
\newcommand{\twp}{{\rm 2p}}
\newcommand{\thp}{{\rm 3p}}
\newcommand{\deta}{\Delta\eta}
\newcommand{\dphi}{\Delta\phi}
\newcommand{\cme}{{\textsc{cme}}}
\newcommand{\ampt}{{\textsc{ampt}}}
\newcommand{\hijing}{{\textsc{hijing}}}
\newcommand{\poi}{{\textsc{poi}}}
\newcommand{\snn}{\sqrt{s_{{\textsc{nn}}}}}
\newcommand{\srp}{{\textsc{rp}}}
\newcommand{\ssp}{{\textsc{sp}}}
\newcommand{\spp}{{\textsc{pp}}}
\newcommand{\sep}{{\textsc{ep}}}
\newcommand{\dg}{\Delta\gamma}
\newcommand{\enf}{\epsilon_{\rm nf}}
\newcommand{\fcme}{f_{\textsc{cme}}}
\newcommand{\minv}{m_{\rm inv}}
\newcommand{\bkgd}{\text{bkgd}}
\newcommand{\signal}{\text{signal}}
\newcommand {\mean}[1]   {\langle{#1}\rangle}

\newcommand {\red}[1]   {\textcolor{red}{#1}}
\newcommand {\blue}[1]  {\textcolor{blue}{#1}}
\newcommand {\green}[1] {\textcolor{green}{#1}}
\newcommand {\gray}[1] {\textcolor{gray}{#1}}
\newcommand {\orange}[1] {\textcolor{orange}{#1}}

%------------------------------------------------------------------------------------------------%

%\linespread{1.6}
\title{
%Estimate of the nonflow background in the isobar measurement Ru+Ru/Zr+Zr ratio of $N\Delta\gamma/v_{2}$ at STAR
%Estimate of nonflow backgrounds in isobar collisions searching for the chiral magnetic effect at the Relativistic Heavy-Ion Collider
Upper Limit on the Chiral Magnetic Effect in Isobar Collisions at the Relativistic Heavy-Ion Collider
}
\affiliation{Abilene Christian University, Abilene, Texas   79699}
\affiliation{Alikhanov Institute for Theoretical and Experimental Physics NRC "Kurchatov Institute", Moscow 117218}
\affiliation{Argonne National Laboratory, Argonne, Illinois 60439}
\affiliation{American University in Cairo, New Cairo 11835, Egypt}
\affiliation{Ball State University, Muncie, Indiana, 47306}
\affiliation{Brookhaven National Laboratory, Upton, New York 11973}
\affiliation{University of Calabria \& INFN-Cosenza, Rende 87036, Italy}
\affiliation{University of California, Berkeley, California 94720}
\affiliation{University of California, Davis, California 95616}
\affiliation{University of California, Los Angeles, California 90095}
\affiliation{University of California, Riverside, California 92521}
\affiliation{Central China Normal University, Wuhan, Hubei 430079 }
\affiliation{University of Illinois at Chicago, Chicago, Illinois 60607}
\affiliation{Chongqing University, Chongqing, 401331}
\affiliation{Creighton University, Omaha, Nebraska 68178}
\affiliation{Czech Technical University in Prague, FNSPE, Prague 115 19, Czech Republic}
\affiliation{National Institute of Technology Durgapur, Durgapur - 713209, India}
\affiliation{ELTE E\"otv\"os Lor\'and University, Budapest, Hungary H-1117}
\affiliation{Frankfurt Institute for Advanced Studies FIAS, Frankfurt 60438, Germany}
\affiliation{Fudan University, Shanghai, 200433 }
\affiliation{Guangxi Normal University, Guilin, 541004}
\affiliation{University of Heidelberg, Heidelberg 69120, Germany }
\affiliation{University of Houston, Houston, Texas 77204}
\affiliation{Huzhou University, Huzhou, Zhejiang  313000}
\affiliation{Indian Institute of Science Education and Research (IISER), Berhampur 760010 , India}
\affiliation{Indian Institute of Science Education and Research (IISER) Tirupati, Tirupati 517507, India}
\affiliation{Indian Institute Technology, Patna, Bihar 801106, India}
\affiliation{Indiana University, Bloomington, Indiana 47408}
\affiliation{Institute of Modern Physics, Chinese Academy of Sciences, Lanzhou, Gansu 730000 }
\affiliation{University of Jammu, Jammu 180001, India}
\affiliation{Joint Institute for Nuclear Research, Dubna 141 980}
\affiliation{Kent State University, Kent, Ohio 44242}
\affiliation{University of Kentucky, Lexington, Kentucky 40506-0055}
\affiliation{Lawrence Berkeley National Laboratory, Berkeley, California 94720}
\affiliation{Lehigh University, Bethlehem, Pennsylvania 18015}
\affiliation{Max-Planck-Institut f\"ur Physik, Munich 80805, Germany}
\affiliation{Michigan State University, East Lansing, Michigan 48824}
\affiliation{National Research Nuclear University MEPhI, Moscow 115409}
\affiliation{National Institute of Science Education and Research, HBNI, Jatni 752050, India}
\affiliation{National Cheng Kung University, Tainan 70101 }
\affiliation{The Ohio State University, Columbus, Ohio 43210}
\affiliation{Panjab University, Chandigarh 160014, India}
\affiliation{NRC "Kurchatov Institute", Institute of High Energy Physics, Protvino 142281}
\affiliation{Purdue University, West Lafayette, Indiana 47907}
\affiliation{Rice University, Houston, Texas 77251}
\affiliation{Rutgers University, Piscataway, New Jersey 08854}
\affiliation{University of Science and Technology of China, Hefei, Anhui 230026}
\affiliation{South China Normal University, Guangzhou, Guangdong 510631}
\affiliation{Sejong University, Seoul, 05006, South Korea}
\affiliation{Shandong University, Qingdao, Shandong 266237}
\affiliation{Shanghai Institute of Applied Physics, Chinese Academy of Sciences, Shanghai 201800}
\affiliation{Southern Connecticut State University, New Haven, Connecticut 06515}
\affiliation{State University of New York, Stony Brook, New York 11794}
\affiliation{Instituto de Alta Investigaci\'on, Universidad de Tarapac\'a, Arica 1000000, Chile}
\affiliation{Temple University, Philadelphia, Pennsylvania 19122}
\affiliation{Texas A\&M University, College Station, Texas 77843}
\affiliation{University of Texas, Austin, Texas 78712}
\affiliation{Tsinghua University, Beijing 100084}
\affiliation{University of Tsukuba, Tsukuba, Ibaraki 305-8571, Japan}
\affiliation{University of Chinese Academy of Sciences, Beijing, 101408}
\affiliation{Valparaiso University, Valparaiso, Indiana 46383}
\affiliation{Variable Energy Cyclotron Centre, Kolkata 700064, India}
\affiliation{Wayne State University, Detroit, Michigan 48201}
\affiliation{Wuhan University of Science and Technology, Wuhan, Hubei 430065}
\affiliation{Yale University, New Haven, Connecticut 06520}

\author{M.~I.~Abdulhamid}\affiliation{American University in Cairo, New Cairo 11835, Egypt}
\author{B.~E.~Aboona}\affiliation{Texas A\&M University, College Station, Texas 77843}
\author{J.~Adam}\affiliation{Czech Technical University in Prague, FNSPE, Prague 115 19, Czech Republic}
\author{J.~R.~Adams}\affiliation{The Ohio State University, Columbus, Ohio 43210}
\author{G.~Agakishiev}\affiliation{Joint Institute for Nuclear Research, Dubna 141 980}
\author{I.~Aggarwal}\affiliation{Panjab University, Chandigarh 160014, India}
\author{M.~M.~Aggarwal}\affiliation{Panjab University, Chandigarh 160014, India}
\author{Z.~Ahammed}\affiliation{Variable Energy Cyclotron Centre, Kolkata 700064, India}
\author{A.~Aitbaev}\affiliation{Joint Institute for Nuclear Research, Dubna 141 980}
\author{I.~Alekseev}\affiliation{Alikhanov Institute for Theoretical and Experimental Physics NRC "Kurchatov Institute", Moscow 117218}\affiliation{National Research Nuclear University MEPhI, Moscow 115409}
% ...

\collaboration{STAR Collaboration}\noaffiliation

%\author{The STAR Collaboration}
%\author{Yicheng Feng}
%\email{feng216@purdue.edu}
%\address{Department of Physics and Astronomy, Purdue University, West Lafayette, IN 47907, USA}
%\author{Fuqiang Wang}
%\email{fqwang@purdue.edu}
%\address{Department of Physics and Astronomy, Purdue University, West Lafayette, IN 47907, USA}
%\author{CME Focus Group}
\date{\today} %this is useful in drafting stage
%\draftversion{12}

%------------------------------------------------------------------------------------------------%

\begin{abstract} 
The chiral magnetic effect (CME) is a phenomenon that arises from the QCD anomaly in the presence of an external magnetic field. The experimental search for its evidence has been one of the key goals of the physics program of the Relativistic Heavy-Ion Collider.
The STAR collaboration has previously presented the results of a blind analysis of isobar collisions (${^{96}_{44}\text{Ru}}+{^{96}_{44}\text{Ru}}$, ${^{96}_{40}\text{Zr}}+{^{96}_{40}\text{Zr}}$) in the search for the CME.
The isobar ratio ($Y$) of CME-sensitive observable, charge separation scaled by elliptic anisotropy, is close to but systematically larger than the inverse multiplicity ratio, the naive background baseline. 
This indicates the potential existence of a CME signal and the presence of remaining nonflow background due to two- and three-particle correlations which are different between the isobars.
In this post-blind analysis, we estimate the contributions from those nonflow correlations as a background baseline to $Y$, utilizing the isobar data as well as \hijing\ simulations.
This baseline is found consistent with the isobar ratio measurement, and an upper limit of 10\% at 95\% confidence level is extracted for the CME fraction in the charge separation measurement in isobar collisions at $\sqrt{s_{\textsc{nn}}}=200$ GeV. 
\end{abstract}

%\pacs{25.75.-q, 25.75.-Gz, 25.75.-Ld} 

%------------------------------------------------------------------------------------------------%

\maketitle

%------------------------------------------------------------------------------------------------%
%\section{Introduction}
{\em Introduction.} The chiral magnetic effect (CME) refers to an electric current (charge separation of produced particles) along the strong magnetic field produced in relativistic heavy-ion collisions due to chirality-imbalanced, parity and charge-parity odd metastable domains~\cite{Fukushima:2008xe}. The formation of such domains has been predicted by quantum chromodynamics (QCD) to occur at high temperatures in those collisions because of vacuum fluctuations~\cite{Morley:1983wr,Kharzeev:1998kz,Kharzeev:2004ey,Kharzeev:2007jp} and may be pertinent to the matter-antimatter asymmetry of our universe~\cite{RevModPhys.76.1}. 

To measure charge separation, three-point correlators, 
\begin{equation} \label{eq:dg}
\begin{split}
	\gamma_{\alpha\beta}=&\mean{\cos(\phi_\alpha+\phi_\beta-2\psi_\srp)} \\
	\dg=&\gamma_\pos-\gamma_\pss,
\end{split}
\end{equation} 
are used ~\cite{Voloshin:2004vk}. 
The terms $\phi_{\alpha,\beta}$ represent the azimuthal angles (in the plane perpendicular to the beam axis) of particles of interest ($\alpha,\beta$), which are of either opposite sign (\pos) or same sign (\pss) in electric charge.
The average $\langle\cdots\rangle$ is taken over particle pairs and events.
%The term $\psi_\srp$ is the azimuthal angle of the reaction plane, defined by the beam and impact parameter directions. %, and is often surrogated by the second azimuthal harmonic plane $\psi_2$ reconstructed by final-state particle distributions.
%The correlators can also be calculated w.r.t.~other reference planes as proxies of $\psi_{\srp}$: 
%the spectator plane $\psi_{\ssp}$ is determined by the spectator nucleons of the collision, among which the protons create the magnetic field and therefore define the magnetic field plane; 
%the second harmonic symmetry plane $\psi_{2}$ is reconstructed by the anisotropy of final-state particles~\cite{Poskanzer:1998yz}, which also fluctuates around $\psi_{\srp}$. 
The azimuthal angle of the reaction plane $\psi_\srp$, defined by the beam and impact parameter directions, is used here because the magnetic field direction fluctuates about $\psi_\srp$. The $\psi_\srp$ is only a theoretical concept, experimentally unmeasurable. It can be approximately determined by the spectator plane $\psi_\ssp$, measured by the spectator neutrons which is more pertinent to the direction of the magnetic field, mainly produced by the spectator protons. The $\psi_\srp$ is often surrogated by the second azimuthal harmonic plane $\psi_2$ reconstructed by final-state particle distributions~\cite{Poskanzer:1998yz}, which fluctuates about $\psi_\srp$.
While charge-independent backgrounds are canceled in $\dg$, backgrounds remain from two-particle (2p) correlations coupled with the elliptic flow of those correlation sources, such as resonances and jets~\cite{Voloshin:2004vk,Wang:2009kd,Bzdak:2009fc,Schlichting:2010qia}.
These backgrounds dominate charge separation measurements at the Relativistic Heavy-Ion Collider (RHIC)~\cite{Abelev:2009ac,Abelev:2009ad,Abelev:2012pa,Adamczyk:2013hsi,Adamczyk:2013kcb,Adamczyk:2014mzf,STAR:2019xzd,STAR:2020gky,STAR:2021pwb} and the Large Hadron Collider (LHC)~\cite{Khachatryan:2016got,Sirunyan:2017quh,Acharya:2017fau,ALICE:2020siw}. 

To eliminate backgrounds, isobar ${^{96}_{44}\text{Ru}}+{^{96}_{44}\text{Ru}}$ and ${^{96}_{40}\text{Zr}}+{^{96}_{40}\text{Zr}}$ collisions at nucleon-nucleon center-of-mass energy of $\snn=200$~GeV were conducted in a single year data collection (2018) by the Solenoid Tracker At RHIC (STAR)~\cite{STAR:2021mii}. Due to the identical mass number, backgrounds were expected to be equal in those collision systems, whereas an appreciable CME signal difference would exist because of the different atomic numbers responsible for the magnetic field~\cite{Voloshin:2010ut,Skokov:2016yrj}. 
However, contrary to expectations, the isobar data~\cite{STAR:2021mii} show that the two systems have different background contributions: the two isobars differ by up to a few percent in the produced charged particle multiplicities ($4.4\%$) and the elliptic flows ($1.4\%$). These differences are consistent with energy density functional calculations of the nuclear structures, resulting in a smaller Ru nucleus than the Zr nucleus~\cite{Xu:2017zcn,Li:2018oec,Xu:2021vpn}. 
%The multiplicity difference resulted in the isobar (Ru+Ru/Zr+Zr) ratio of the quantity $\dg/v_2$ smaller than unity. 
Although the $\dg/v_{2}$ was constructed to account for the elliptic flow (parameterized by $v_{2}$) difference, the isobar (Ru+Ru/Zr+Zr) ratio of the $\dg/v_{2}$ measurements, $Y \equiv \frac{(\Delta\gamma/v_{2}^{*})^{\Ru}}{(\Delta\gamma/v_{2}^{*})^{\Zr}}$, was smaller than unity due to the multiplicity difference that was not considered in the blind analysis~\cite{STAR:2021mii}.
%If number of correlation sources is proportional to multiplicity, then isobar ratio of $\dg/v_2$ would equal to that of the inverse multiplicity ($N$) for pure background; the data show it is larger~\cite{STAR:2021mii}, suggesting finite CME signal~\cite{Kharzeev:2022hqz}.
 
If the number of correlation sources is proportional to multiplicity, then $Y$ would be equal to the isobar ratio of the inverse multiplicity ($1/N$) for a pure flow-driven background scenario. A quantitative comparison shows that $Y$ is slightly larger than the $1/N$ ratio~\cite{STAR:2021mii}, indicating the potential presence of a CME signal~\cite{Kharzeev:2022hqz}.
However, the measurement of the relative pair excess $r=(N_\pos-N_\pss)/N_\pos$~\cite{STAR:2021mii} indicates a violation of such proportionality, 
which is one indication that this naive baseline is not strictly correct.  
%The isobar ratio of $\dg/v_2$ is smaller than that of $r$~\cite{STAR:2021mii}, suggesting a ``negative'' CME. 
%Obviously, a more rigorous assessment of the background baseline is called for, which is the subject of this Letter. 
In order to search for any residual signals of CME, a more rigorous evaluation of the background baseline is necessary, which is the main goal of this Letter.
Further details of the background assessment analysis can be found in the long companion paper~\cite{STAR:2023ioo}.

\begin{figure*}
    \includegraphics[width=0.325\linewidth]{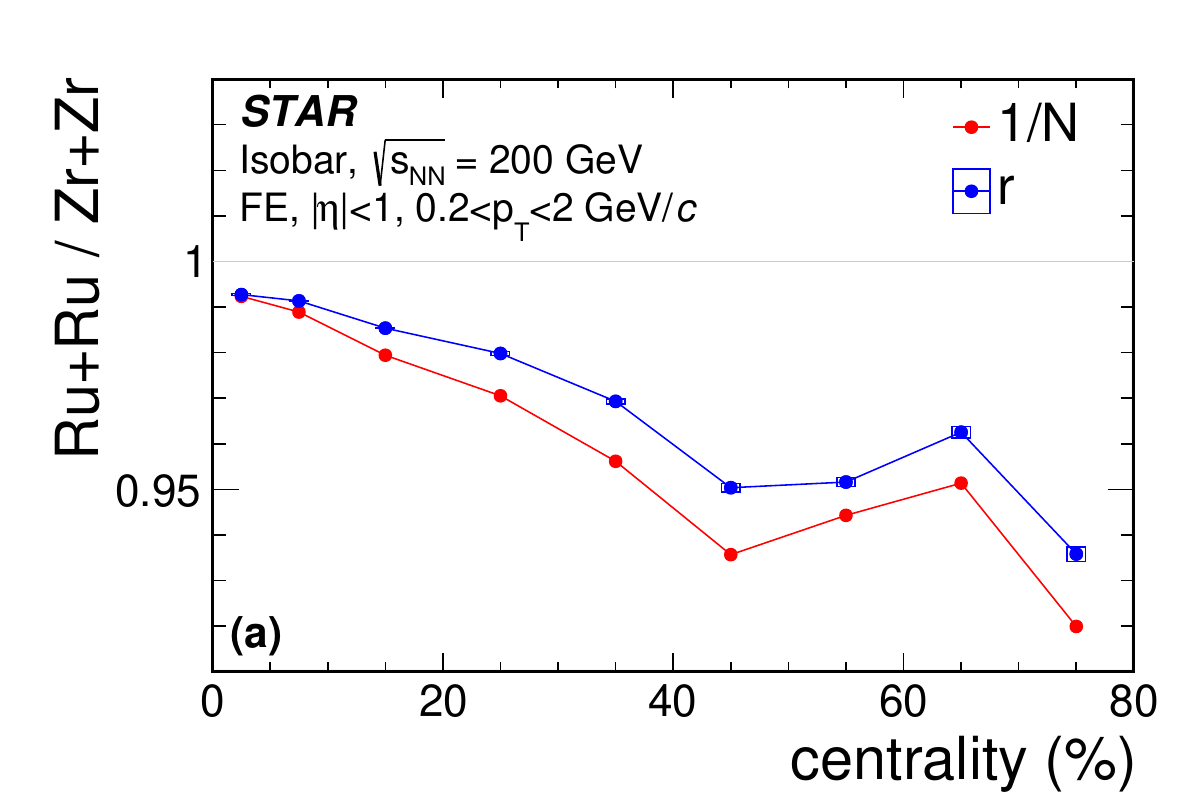}
    \includegraphics[width=0.325\linewidth]{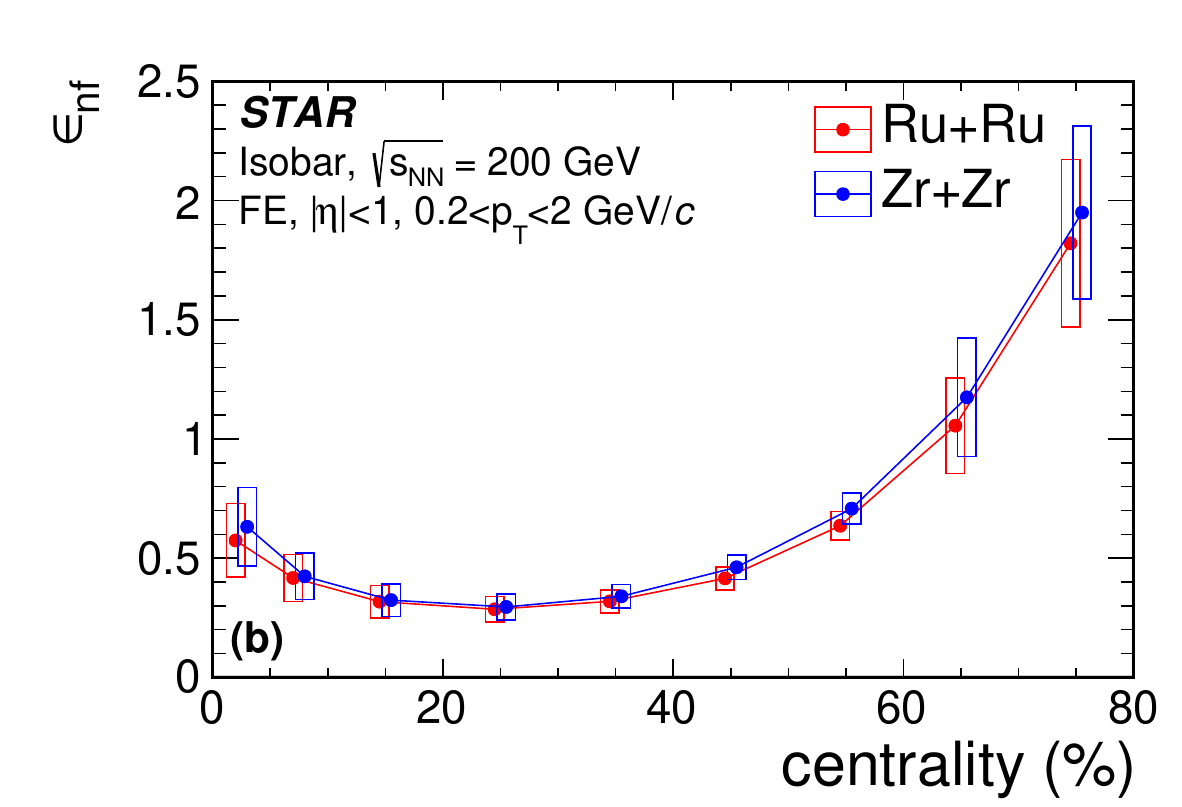}
    \includegraphics[width=0.325\linewidth]{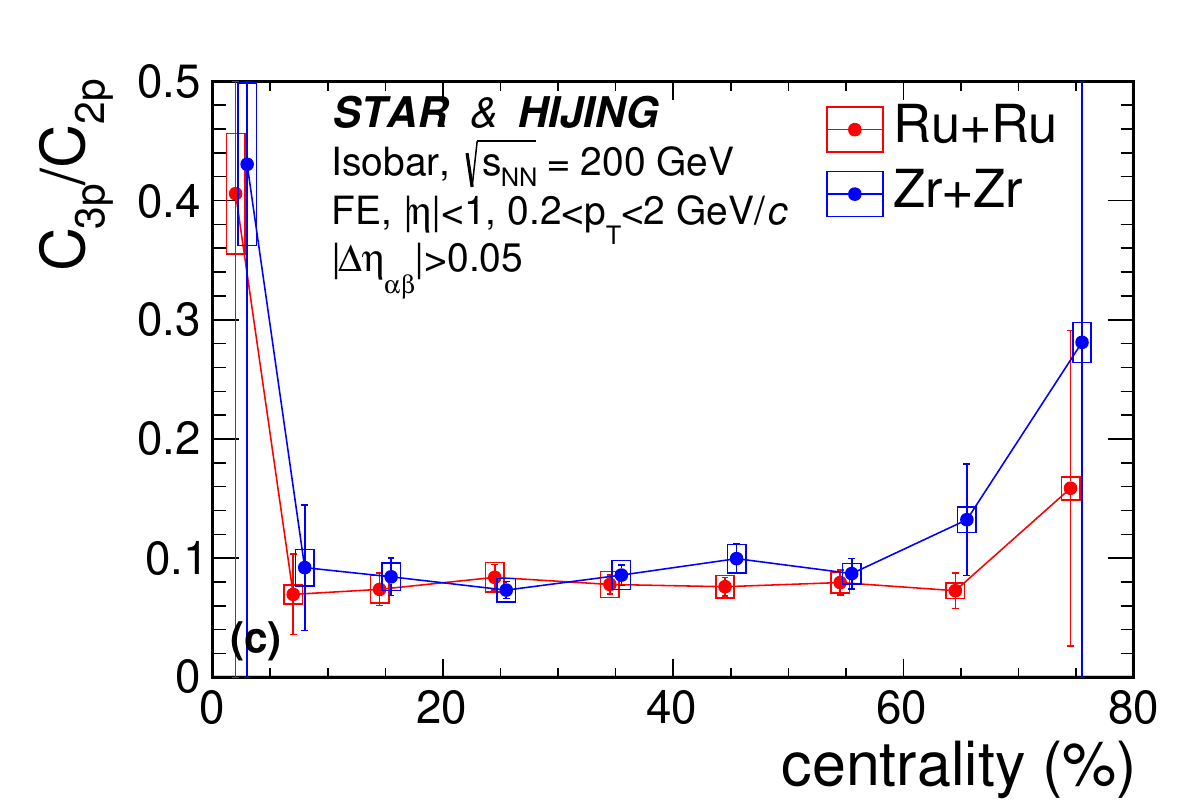}
    %\includegraphics[width=0.325\linewidth]{FulEnf}
    %\includegraphics[width=0.49\linewidth]{SubEnf}
    %\caption{$\epsilon_{\text{nf}}$}
    %\includegraphics[width=0.325\linewidth]{FulE3E2}
    %\includegraphics[width=0.49\linewidth]{SubE3E2}
    %\caption{$\epsilon_{3}/\epsilon_{2}$}
    \caption{ 
	%Isobar ratios of quantities for full-event (FE) with Group-3 cuts~\cite{STAR:2021mii}, others are similar. 
	%(a) $r\equiv(N_\pos-N_\pss)/N_\pos$ and inverse multiplicity ($N$) ratios; (b) nonflow $v_2$ contamination $\enf$; (c) $C_\thp/C_\twp$ where $C_\thp$ is estimated by \hijing\ and $C_\twp$ is from ZDC measurement in~\cite{STAR:2021mii}.
	(a) isobar ratio of $r\equiv(N_\pos-N_\pss)/N_\pos$ and inverse multiplicity ($1/N$); (b) nonflow $v_2$ contamination $\enf$; (c) $C_\thp/C_\twp$ where $C_\thp$ is estimated using \hijing\ and $C_\twp$ is from ZDC measurement in~\cite{STAR:2021mii}. All quantities in these plots use Group-3 full-event (FE) cuts from~\cite{STAR:2021mii}; other cuts give similar results. 
	}
    \label{fig}
\end{figure*}
%------------------------------------------------------------------------------------------------%

{\em Refined baseline.}
In off-center heavy-ion collisions, the azimuthal distribution of final-state particles is anisotropic because of the anisotropic expansion of the collision fireball~\cite{Ollitrault:1992bk}. The second azimuthal harnomic, elliptic flow, is used to reconstruct $\psi_\srp$, the accuracy of which is corrected by a resolution factor~\cite{Poskanzer:1998yz}. Equivalent to Eq.~(\ref{eq:dg}) with $\psi_{2}$, but more directly, $\gamma$ can also be calculated using the three-particle (3p) correlator~\cite{Abelev:2009ad}, $C_{3,\alpha\beta}=\mean{\cos(\phi_\alpha+\phi_\beta-2\phi_c)}$ and $\gamma_{\alpha\beta}=C_{3,\alpha\beta}/v_2$, where $v_2$ is the elliptic flow of particles of type $c$, which are usually taken as all charged hadrons in a given detector acceptance. %The background contribution can be expressed~\cite{Feng:2021pgf} as
The background contribution to $\Delta\gamma/v_{2}$ from intrinsic 2p and 3p correlation can be expressed~\cite{Feng:2021pgf} as
\begin{equation}
%    \Delta\gamma_{\rm bkgd}=\frac{N_\twp}{N_{\alpha}N_{\beta}}\langle\cos(\phi_{\alpha}+\phi_{\beta}-2\phi_\twp)\rangle v_{2,\twp}\,.
    \frac{\dg_{\bkgd}}{v_2^*}=\frac{C_\twp}{N}\frac{v_2^2}{v_2^{*2}}+\frac{C_\thp}{N}\frac{1}{N_cv_2^{*2}}
    =\frac{C_\twp}{N}\frac{1+\frac{C_\thp/C_\twp}{Nv_2^2}}{1+\epsilon_{\rm nf}}\,,
    \label{eq:bkgd}
\end{equation} 
where
\begin{eqnarray}
    \frac{C_\twp}{N}&=&\frac{N_\twp}{N_\pos}\left(C_{\twp,\pos}\frac{v_{2,\twp}}{v_2}-\frac{\gamma_\pss}{v_2}\right)\,,
    \label{eq:c2p}\\
    \frac{C_\thp}{N}&=&\frac{N_{\thp,\pos}}{N_\pos}C_{\thp,\pos}-\frac{N_{\thp,\pss}}{N_\pss}C_{\thp,\pss}\,.
	\label{eq:c3p}
\end{eqnarray}
%and %the shorthand notations 
The notation $C_{\twp,\pos}=\mean{\cos(\phi_{\alpha}+\phi_{\beta}-2\phi_\twp)}_{\twp,\pos}$ refers to those correlated background pairs only, where $\phi_\twp$ is the azimuth of the pair, $N_\pos$ and $N_\pss$ are \pos\ and \pss\ pair multiplicities, and $N_\twp\equiv N_\pos-N_\pss$. Similarly,  
$C_{\thp,\pos}=\mean{\cos(\phi_{\alpha}+\phi_{\beta}-2\phi_c)}_{\thp,\pos}$ and $C_{\thp,\pss}$ refer to those correlated background triplets only, where $N_{\thp,\pos}$ and $N_{\thp,\pss}$ are their triplet multiplicities.
$N$ is the multiplicity of particles of interest (POI), and $N_c$ is that of particle $c$ (in this analysis $N=N_c$). 
%$N_\pos$ and $N_\pss$ are numbers of \pos\ and \pss\ pairs, $N_\twp\equiv N_\pos-N_\pss$ is number of correlated \pos\ pair  excess and $\phi_\twp$ refers to their azimuths; $N_{\thp,\pos}$ and $N_{\thp,\pss}$ are numbers of correlated \pos-$c$ and \pss-$c$ triplets.
The $v_2$ and $v_{\rm 2,2p}$ refer to the {\em true} elliptic flow of POIs and those correlated 2p sources, respectively. The quantity $v_2^*$ refers to the measured elliptic flow, which contains {\em nonflow}--correlations unrelated to the global collision geometry. 
The equation, 
\begin{equation} \label{eq:enf}
    \epsilon_{\rm nf}=(v_2^*/v_2)^2-1 
    \,,
\end{equation}
quantifies the relative nonflow contamination.

The isobar ratio can then be decomposed into 
\begin{widetext}
\begin{equation}
    Y_{\bkgd} \equiv\frac{\left(\dg_{\bkgd}/v_2^*\right)^{\Ru}}{\left(\dg_{\bkgd}/v_2^*\right)^{\Zr}}
    \approx1+\frac{\delta(C_\twp/N)}{C_\twp/N}-\frac{\delta\enf}{1+\enf}+\frac{1}{1+\frac{Nv_2^2}{C_\thp/C_\twp}}\left(\frac{\delta C_\thp}{C_\thp}-\frac{\delta C_\twp}{C_\twp}-\frac{\delta N}{N}-\frac{\delta v_2^2}{v_2^2}\right)\,,
    \label{eq:Y}
\end{equation}
\end{widetext}
%where $\delta X\equiv X^{\Ru}-X^{\Zr}$ and all other quantities refer to those in Zr+Zr. 
where $\delta X\equiv X^{\Ru}-X^{\Zr}$ for any $X=C_{\thp}$, $C_{\twp}$, etc., while all other quantities without ``$\delta$'' refer to those in Zr+Zr.
%As is evident from Eq.~(\ref{eq:Y}), the nonflow contamination in $v_2$ and intrinsic three-particle (3p) correlations affect the background baseline $Y_{\bkgd}$.
Equation~(\ref{eq:Y}) suggests categorizing the nonflow contributions to the background into three ingredients: (1) $\delta(C_{\twp}/N)/(C_{\twp}/N)$ which characterizes the relative difference of flowing clusters between the two isobars; (2) differences that arise from using $v_{2}^{*}$ rather than true flow in the calculation of $\Delta\gamma$, characterized by $\enf$; (3) differences in the relative amounts (or character) of three particle clusters between the isobars.  In the next section we will discuss each of these three in turn.
We note that global spin alignment of $\rho$ mesons can introduce an additional background to the CME~\cite{Shen:2022gtl}. Effect of such a background on isobar measurements needs to be assessed in future studies. 

\begin{figure}
	\includegraphics[width=1.0\linewidth]{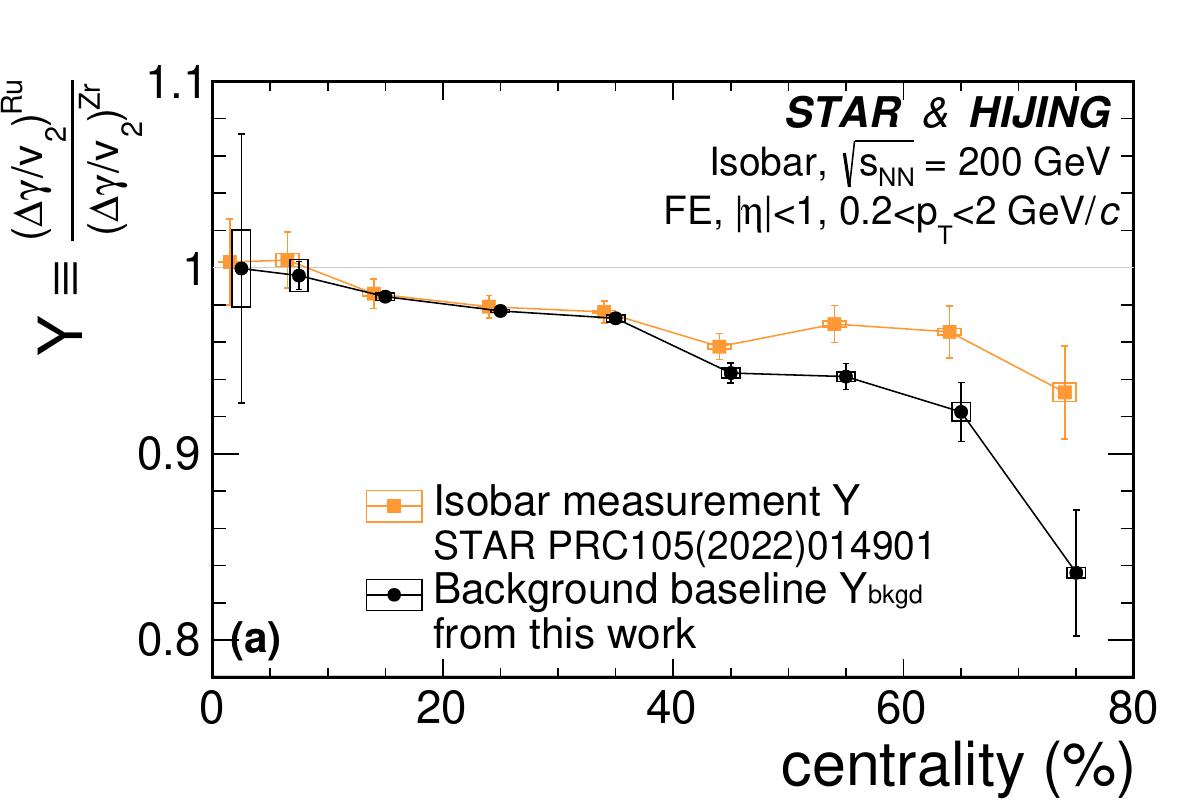}
	\includegraphics[width=1.0\linewidth]{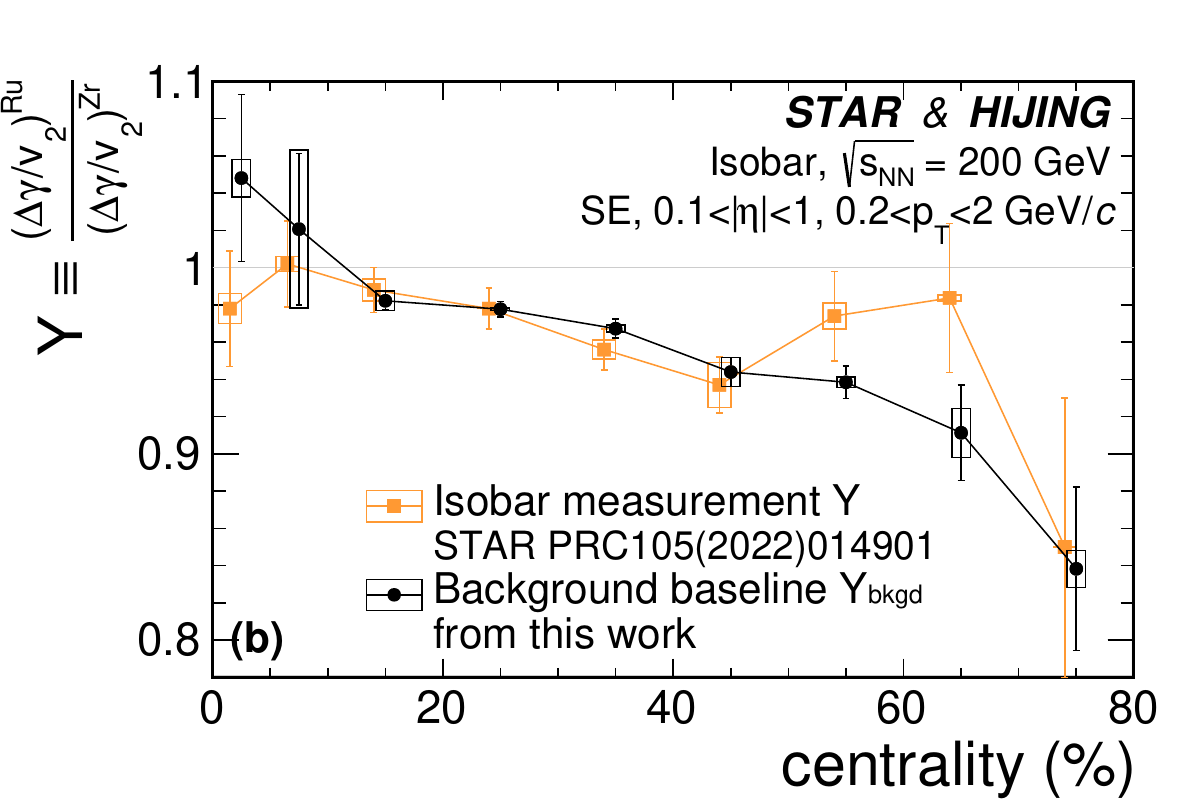}
	\caption{Estimate of background baseline $Y_{\bkgd}$ for the isobar measurement $Y=\frac{(\dg/v_2)^{\Ru}}{(\dg/v_2)^{\Zr}}$ as a function of centrality for (a) the full-event (FE) analysis of Group-3 and (b) the subevent (SE) analysis of Group-2; others are similar.}
	\label{fig:baseline}
\end{figure}

{\em Analysis.} 
%Seven $\dg/v_2$ measurements were published~\cite{STAR:2021mii}; four used two-particle cumulant for the $v_2$ measurement and three-particle correlator for $\dg$; 
The isobar blind analysis~\cite{STAR:2021mii} presented seven different measurements of $\dg / v_{2}$ from four groups; four of these measurements utilized the 2p cumulants for the $v_{2}$ measurement and the 3p correlators for $\dg$.
%The blind analysis~\cite{STAR:2021mii} had seven $\Delta\gamma/v_2$ measurements from four groups. We will use four of them from Group-2,3,4, which used two-particle cumulants for $v_2$ and three-particle correlators for $\Delta\gamma$.
The other three employed the event-plane method, which is similar, but the nonflow effects are more complicated to assess. 
%Several research ``Groups'' participated in the isobar blind analysis~\cite{STAR:2021mii}, and each of them contributed one or more of these measurements.
We focus on the four cumulant measurements with their corresponding analysis cuts with subtle differences. 
%We focus on the four measurements using cumulant methods that only differ by slight variations of analysis cuts.
The same event selections and track quality cuts are used as those in the isobar blind analysis~\cite{STAR:2021mii}. 
%These include, among others, the primary vertex range in $-35\sim25$~cm longitudinal from the center of the STAR Time Projection Chamber (TPC), the track's distance of closest approach from the primary vertex within 3~cm, and requiring the number of fit points to be greater than 15. 
%\orange{[other cuts? $V_{\perp}<2$ cm, $|V_{z} - V_{z}^{\text{VPD}}|<5$ cm, $0.2 < \pt < 2.0$ \gevc, $|\eta|<1$, subevent $0.1 < |\eta| < 1$.]}

The background baseline estimate of Eq.~(\ref{eq:Y}) requires three ingredients. The first ingredient $\delta(C_\twp/N)/(C_\twp/N)$, which is related to 2p nonflow, is primarily determined by $N_\twp/N_\pos$, since $(C_{\twp,\pos}v_{2,\twp}-\gamma_\pss)/v_2$ (dominated by the first term) 
should be highly similar between the isobar systems. We analyze $r\equiv N_\twp/N_\pos$ of identified pions as done in~\cite{STAR:2021mii}.
Since $(C_{\twp,\pos}v_{2,\twp}-\gamma_\pss)/v_2$ likely depends on the pair invariant mass ($\minv$), we take the average $\delta r/r$ over the entire $\minv$ range as the default and assess systematic uncertainties by considering the range $m_{\rm inv}<1$~GeV/$c^2$~\cite{STAR:2023ioo}. Figure~\ref{fig}(a) shows the isobar ratio of $r$ as a function of centrality from the full-event analysis. For comparison, the efficiency-corrected inverse POI multiplicity ratio is also shown. 
The $\delta (C_\twp / N) / (C_\twp / N) \approx\delta r/r$ value averaged over 20--50\% centrality is on the order of $-3\%$~\cite{STAR:2023ioo}. Consequently, the baseline $Y_{\bkgd}$ is altered by this amount from unity.
%The $\delta r/r$ values averaged over 20--50\% centrality are tabulated in Table~\ref{tab} with the without cut on $\deta_{\alpha\beta}$.

The second ingredient is the nonflow contamination in the $v_2^*$ measurement. To estimate it, we fit the acceptance-corrected $(\deta,\dphi)$ 2p correlations for \pss\ pairs from the full-event analysis by $\deta$-independent flow harmonics plus $\deta$- and $\dphi$-dependent nonflow contributions~\cite{STAR:2023ioo}. 
The true flow is assumed to be the same for the \pos\ and \pss\ pairs.
The fitted $v_2$ parameter, as an estimate of true $v_{2}$, is approximately 5.5\% in the 20--50\% centrality range, with a relative difference of approximately 2.2\% between Ru+Ru and Zr+Zr collisions~\cite{STAR:2023ioo}.
With the fitted $v_2$, the $\enf$ can be readily calculated from the $v_2^*$ cumulant measurements~\cite{STAR:2021mii}. %it can also be calculated from the full-event and subevent $(\deta,\dphi)$ correlations in this work and is found to be consistent. 
The $v_2^*$ measurements used slightly different $\deta$ gaps and various methods; the full-event $v_2^*$ from Group-2 applied Gaussian fits in $\deta$ to reduce short-range nonflow contributions~\cite{STAR:2021mii}.
The $\enf$ value ranges from 18--34\% depending on the analysis methods. %is on the order of 30\% and 15\% for full event and subevent analysis, resepctively. 
The systematic uncertainties on $\enf$ are estimated by applying a different acceptance-correction method for the $(\deta,\dphi)$ correlations~\cite{STAR:2023ioo}, by comparing the calculated $v_2^*$ from this analysis to those measured in~\cite{STAR:2021mii}, and by the observed 3\% flow decorrelation over one unit of pseudorapidity based on a separate study from STAR~\cite{Yan:2022qm}. 
%\red{However, any overall systematic uncertainties on the true $v_2$ %%e.g., those from longitudinal decorrelation and charge dependence, 
%cancel in $\delta\enf/(1+\enf)$. Systematic uncertainties on $\delta\enf/(1+\enf)$ are estimated by...}
%The last, common to both isobar systems, cancels in $\delta\enf/(1+\enf)$ and only the former two sources contribute.
The last source, common to both isobars, cancels in $\delta\enf /(1 + \enf)$. %, and only the first two matter. 
Figure~\ref{fig}(b) shows $\enf$ as a function of centrality from the full-event analysis without an $\eta$ gap. 
The $\enf$ value is smaller in Ru+Ru than in Zr+Zr because of the larger multiplicity dilution in the former; %, similar to the measurement of another azimuthal correlator $\Delta\delta$~\cite{STAR:2021mii}; 
the actual nonflow correlation strength after factoring out the multiplicity difference is larger in Ru+Ru than Zr+Zr by approximately 2\%.
The $-\delta\enf/(1+\enf)$ value ranges from 0.6\% to 1.5\%, being smaller for subevent than for full event~\cite{STAR:2023ioo}. 
%$Y_{\bkgd}$ is raised by this amount up towards unity.
This correction increases $Y_\bkgd$ by this amount. % towards unity.
%The average $\enf$ values from various analyses methods are listed in Table~\ref{tab}.

The third ingredient is the genuine 3p correlation background. As 3p correlation measurements are challenging to measure due to the substantial combinatorial background in heavy-ion collisions, we resort to \hijing\ (Heavy Ion Jet INteraction Generator) simulations~\cite{Wang:1991hta,Gyulassy:1994ew}. 
Since \hijing\ conserves charge and does not have flow, the inclusive 3p correlation from \hijing\ is in entirety the $C_\thp$, purely from the correlated triplets. 
%This model ensures charge conservation, which makes the charge-dependent quantity $C_{\thp}$ more realistic. 
The $C_\thp$ from \hijing\ simulations with jet quenching is taken as default, and that from quenching-off simulations (about 20\% higher) is considered as one side of the maximum systematic uncertainty (i.e., the quoted uncertainty is $1/\sqrt{3}$ of that) with the other side treated symmetrically~\cite{STAR:2023ioo}.  
The systematic uncertainties on $\delta C_\thp/C_\thp$ are assessed similarly.
\hijing\ is found to give an adequate description of the peripheral data~\cite{STAR:2023ioo},
%we have examined $K_S$- and $\Lambda$-hadron correlations in midcentral collisions and found their contributions to $C_\thp$ to be similar in data and \hijing; these 
suggesting that \hijing\ is a reliable estimator for $C_\thp$. %The difference is well within the $\pm20\%$ systematic uncertainty from quenching-on and -off comparison. 

\begin{figure*}
    \includegraphics[width=1.0\linewidth]{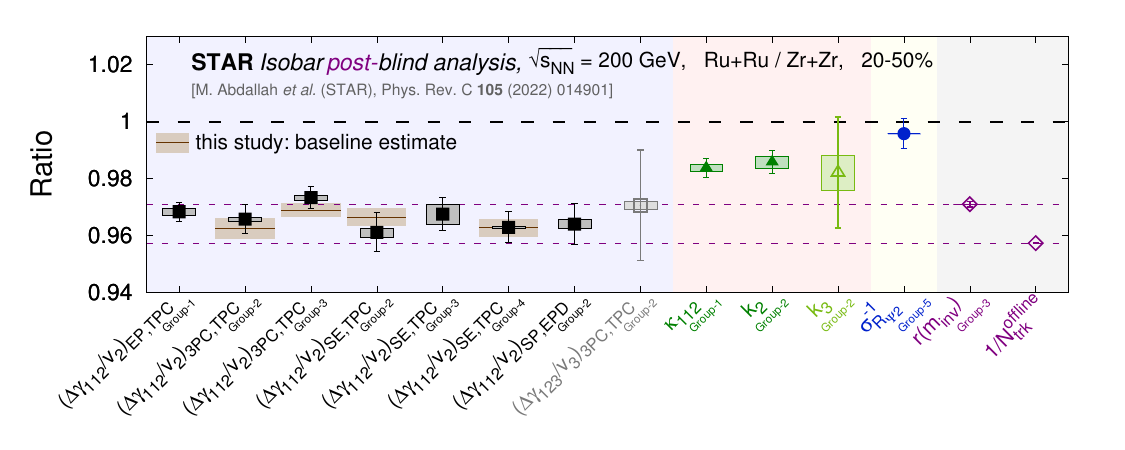}
	\caption{Isobar measurements of $Y$ from the STAR blind analyses~\cite{STAR:2021mii}, together with background baseline estimates for the four measurements that used the cumulant method.}
	\label{fig:moneyplot}
\end{figure*}

The effect of 3p correlation on $Y_{\bkgd}$ also depends on $C_\twp$. The $C_\twp$ value can be estimated directly from the corresponding zero-degree calorimeter (ZDC) measurements of $N\dg/v_2$ because it largely eliminates $v_2$ nonflow and 3p correlations due to the large $\eta$ gap between the ZDC and TPC. 
%Since ZDC detects the spectator neutrons outside beam rapidity, its plane is regarded as a measurement of $\psi_{\ssp}$ with some resolution. 
A one-sided $-5\%$ systematic uncertainty is assigned to account for any possibly small CME signal contained in the measurement.
%, \red{in addition to those from the ZDC measurements themselves.}
No ZDC measurement is available in~\cite{STAR:2021mii} corresponding to the Group-2 subevent analysis cuts, so it is analyzed in this work to estimate $C_{\twp}$. %It is found to be similar to the average $\dg/v_2$ between those from Group-3 (smaller value because of the wider $0.05<|\eta|<1$ acceptance) and Group-4 (larger value because of no $\deta_{\alpha\beta}$ cut)~\cite{STAR:2021mii} is taken to calculate $C_\twp$.
Figure~\ref{fig}(c) shows $C_\thp/C_\twp$ as a function of centrality for the full-event analysis from approximately 7.0 billion \hijing\ events for each isobar. The relative strength of 3p to 2p correlations is on the order of 10\%. 
The average contribution to $Y_{\bkgd}$ from 3p correlation backgrounds in 20--50\% centrality is around $-1.3\%$. %; it is negative and shifts the baseline away from unity.
%The average values over 20--50\% centrality are listed in Table~\ref{tab}.

{\em Results.}
Figure~\ref{fig:baseline} shows the estimated baseline as a function of centrality, along with the isobar data corresponding to the Group-3 full-event and Group-2 subevent analysis, respectively.
The systematic uncertainty on the baseline is taken to be the quadratic sum of the uncertainties on the individual components as described above. 
Figure~\ref{fig:moneyplot} depicts the $Y$ measurements in the 20--50\% centrality range~\cite{STAR:2021mii} together with our estimated baselines averaged over the same range. The three terms in the baseline (Eq.~(\ref{eq:Y})) are averaged over centrality individually and then summed. 
The measured data are consistent with these estimated baselines over most of the centrality bins.

%Four results published in Ref.~\cite{STAR:2021mii} used the two-particle cumulant for $v_2$ and the three-particle correlator for $\dg$ measurements, with slightly different cuts. 
%We tabulate the corresponding background baselines in Table~\ref{tab}.
%Table~\ref{tab} lists the estimated baselines together with isobar data for the four measurements.
%Table~\ref{tab} tabulates the four cumulant measurements of $Y$ and the estimated baselines $Y_{\bkgd}$ in the centrality range of 20--50\%. Also tabulated are their 
The differences, $Y-Y_{\bkgd}$, measure the purported CME signal.
The baseline estimates partly come from data. These estimates are derived from quantities and/or methods different from the $\dg/v_2$ measurements, so their statistical uncertainties are treated independently. 
The systematic uncertainties on the isobar measurements were assessed by varying analysis cuts, and were found to be significantly smaller than the statistical uncertainties~\cite{STAR:2021mii}. We did not repeat those in the baseline calculation ($Y_{\bkgd}$) to avoid double counting in systematics, but instead propagate data systematic uncertainties to differences $Y - Y_{\bkgd}$ in quadrature. 
%The estimated background baselines are approximately \red{1\% and 0.5\%} above unity for the full-event and subevent data, respectively.

\begin{figure}
    \centering
    \includegraphics[width=1.0\linewidth]{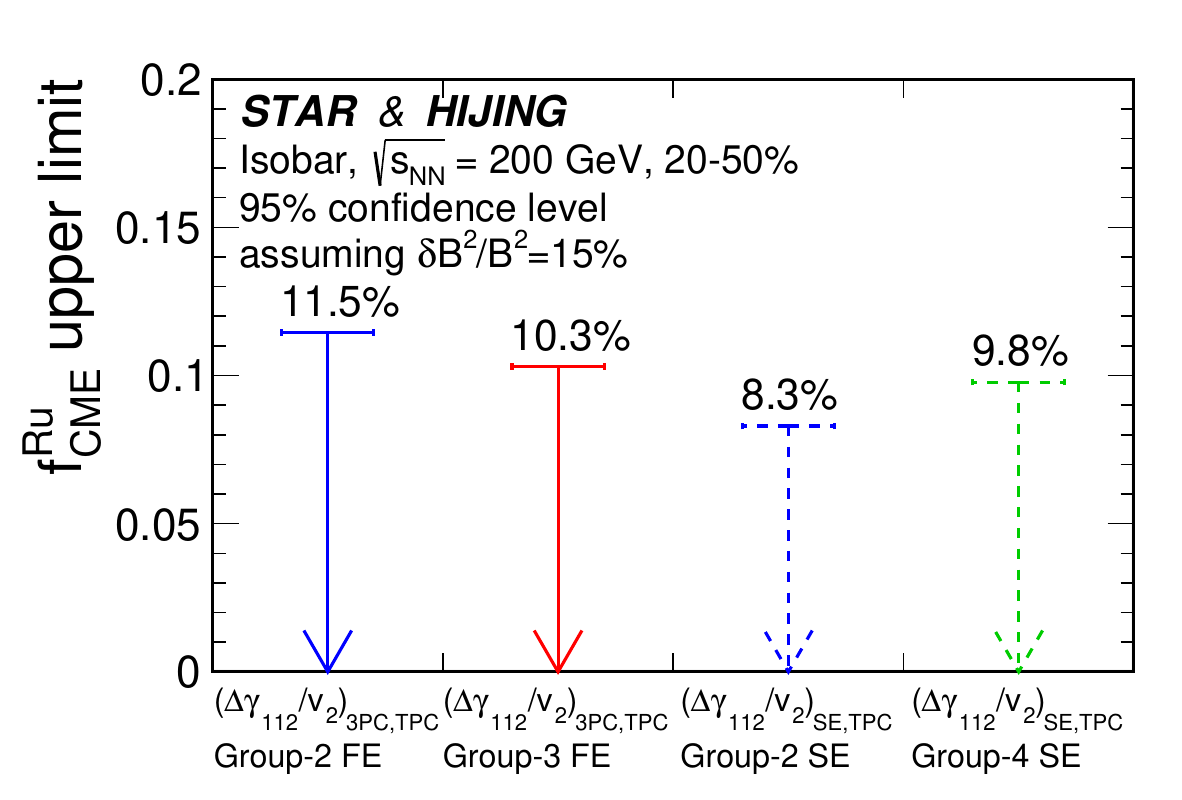}
    \caption{The $\fcme$ upper limits at 95\% confidence level from Ru+Ru collisions for the four results in Fig.~\ref{fig:moneyplot}. }
    \label{fig:limit}
\end{figure}

Our results indicate that the CME signal difference in isobar collisions is consistent with zero within uncertainties. 
We therefore estimate the upper limit of the possible CME signals. 
%Comparing isobar collisions, we are sensitive to the difference in the magnetic field strengths.
The isobar difference in the magnetic field strengths makes it possible to extract CME signal by comparing the isobar collision systems. 
Assuming it results in a  15\% difference in the CME $\dg$ signal~\cite{Skokov:2009qp,Bzdak:2011yy,Deng:2012pc}, then our results translate into 
%a CME signal fraction of \red{$\fcme=(3.3\pm3.1\pm1.5)\%$, or an upper limit of 10\% at 95\% confidence level}.
an accuracy of a few percent on the CME signal fraction ($\fcme$)~\cite{STAR:2023ioo}. 
%Assuming Gaussian with boundary at origin~\cite{Feldman:1997qc} ($\fcme \ge 0$), 
We extract an upper limit of $\fcme \sim 10\%$ at 95\% confidence level~\cite{Feldman:1997qc} for Ru+Ru collisions (Zr+Zr is similar).
Figure~\ref{fig:limit} depicts those upper limits for the four results shown in Fig.~\ref{fig:moneyplot}.

%------------------------------------------------------------------------------------------------%
{\em Summary.}
%We estimate the deviations of the three-particle correlator from the simple inverse multiplicity scaling using pion pair multiplicities in isobar collisions. 
%We extract nonflow contamination in the $v_{2}$ measurements by fitting 2p $(\Delta\eta,\Delta\phi)$ distributions. 
%We evaluate the contributions from genuine three-particle correlations to the three-particle correlator by using \hijing\ simulations. 
%Using these inputs, we arrive at estimates of background baselines for the isobar ${^{96}_{44}\text{Ru}}+{^{96}_{44}\text{Ru}}$ to ${^{96}_{40}\text{Zr}}+{^{96}_{40}\text{Zr}}$ ratio of the $N\dg/v_2$ variable. 
%The estimated baselines are consistent with the STAR measurements.
%We extract an upper limit on the CME fraction at 95\% confidence level to be $\sim 10\%$ of the $\dg$ measurements in isobar collisions at $\snn=200$~GeV at RHIC.
We reexamine the isobar ratio $Y \equiv \frac{(\Delta\gamma/v_{2})^{\Ru} }{ (\Delta\gamma/v_{2})^{\Zr}}$, which measures the charge separation due to the chiral magnetic effect (CME) in isobar collisions, and account for the background effects from multiplicity and nonflow correlations. %s like resonance decays, jets, etc. 
%We use a new method to estimate a background baseline for $Y$ by comparing 2p and 3p correlations from STAR isobar data and HIJING simulations. 
We estimate the background baseline for $Y$ by using 2p and 3p correlations from STAR isobar data and \hijing\ simulations. 
The estimated baselines agree with the STAR measurements~\cite{STAR:2021mii}. 
%We then set a $95\%$ upper-limit on the CME fraction of $\sim 10\%$ in isobar collisions at $\snn = 200$ GeV at RHIC. 
We set an upper limit at the 95\% confidence level on the CME fraction of $\sim 10\%$ in isobar collisions at $\snn = 200$ GeV at RHIC. 
%This study provides a solid interpretation to the previous STAR isobar measurements. %, and develops a workflow to estimate backgrounds in other and future related CME searches.
This study provides a robust interpretation and a highly anticipated estimation of the remaining CME signal in the STAR isobar measurements. 

%------------------------------------------------------------------------------------------------%

\section*{acknowledgments}

We thank the RHIC Operations Group and RCF at BNL, the NERSC Center at LBNL, and the Open Science Grid consortium for providing resources and support.  This work was supported in part by the Office of Nuclear Physics within the U.S. DOE Office of Science, the U.S. National Science Foundation, National Natural Science Foundation of China, Chinese Academy of Science, the Ministry of Science and Technology of China and the Chinese Ministry of Education, the Higher Education Sprout Project by Ministry of Education at NCKU, the National Research Foundation of Korea, Czech Science Foundation and Ministry of Education, Youth and Sports of the Czech Republic, Hungarian National Research, Development and Innovation Office, New National Excellency Programme of the Hungarian Ministry of Human Capacities, Department of Atomic Energy and Department of Science and Technology of the Government of India, the National Science Centre and WUT ID-UB of Poland, the Ministry of Science, Education and Sports of the Republic of Croatia, German Bundesministerium f\"ur Bildung, Wissenschaft, Forschung and Technologie (BMBF), Helmholtz Association, Ministry of Education, Culture, Sports, Science, and Technology (MEXT), Japan Society for the Promotion of Science (JSPS) and Agencia Nacional de Investigaci\'on y Desarrollo (ANID) of Chile.

%------------------------------------------------------------------------------------------------%

\bibliography{./ref}

%------------------------------------------------------------------------------------------------%

\end{document}